\documentclass[runningheads]{llncs}

\usepackage{graphicx}
\usepackage{setspace}
\usepackage{amsmath,amssymb,amsfonts}
\usepackage{algorithmic}
\usepackage{graphicx}
\usepackage{textcomp}
\usepackage{xcolor}
\usepackage{subfig}
\usepackage{mathrsfs}
\usepackage{scalefnt}
\usepackage{multirow}
\usepackage{tcolorbox}
\usepackage{booktabs}
\usepackage{bm}
\usepackage{multirow}
\usepackage[normalem]{ulem}
\usepackage{lipsum}

\begin{document}

\title{LLM is Knowledge Graph Reasoner: LLM’s Intuition-aware Knowledge Graph Reasoning for Cold-start Sequential Recommendation}

\titlerunning{LLM is KG Reasoner}

\author{Keigo Sakurai\inst{1}\and
Ren Togo\inst{2} \and
Takahiro Ogawa\inst{2}  \and
Miki Haseyama\inst{2}}

\authorrunning{K. Sakurai {\it et al}.}

\institute{Graduate School of Information Science and Technology, Hokkaido University, Sapporo Hokkaido 060-0814, Japan\\
\email{sakurai@lmd.ist.hokudai.com}\\\and
Faculty of Information Science and Technology, Hokkaido University, Sapporo Hokkaido 060-0814, Japan\\
\email{\{togo,ogawa,mhaseyama\}@lmd.ist.hokudai.com}}

\maketitle

\begin{abstract}
Knowledge Graphs (KGs) represent relationships between entities in a graph structure and have been widely studied as promising tools for realizing recommendations that consider the accurate content information of items. 
However, traditional KG-based recommendation methods face fundamental challenges: insufficient consideration of temporal information and poor performance in cold-start scenarios.
On the other hand, Large Language Models (LLMs) can be considered databases with a wealth of knowledge learned from the web data, and they have recently gained attention due to their potential application as recommendation systems. 
Although approaches that treat LLMs as recommendation systems can leverage LLMs' high recommendation literacy, their input token limitations make it impractical to consider the entire recommendation domain dataset and result in scalability issues. 
To address these challenges, we propose a LLM's Intuition-aware Knowledge graph Reasoning model (LIKR). 
Our main idea is to treat LLMs as reasoners that output intuitive exploration strategies for KGs.
To integrate the knowledge of LLMs and KGs, we trained a recommendation agent through reinforcement learning using a reward function that integrates different recommendation strategies, including LLM's intuition and KG embeddings. 
By incorporating temporal awareness through prompt engineering and generating textual representations of user preferences from limited interactions, LIKR can improve recommendation performance in cold-start scenarios. 
Furthermore, LIKR can avoid scalability issues by using KGs to represent recommendation domain datasets and limiting the LLM's output to KG exploration strategies. 
Experiments on real-world datasets demonstrate that our model outperforms state-of-the-art recommendation methods in cold-start sequential recommendation scenarios.
\end{abstract}

\section{Introduction}
\label{intro}
Knowledge graphs (KGs) have emerged as promising tools for implementing recommendation systems that consider detailed content information of items~\cite{guo2020survey,ji2021survey,wang2021learning}.
KGs can be considered as databases composed of entities (e.g., users, items, producers) and relations (e.g., ``an item is provided by a producer"), capturing complex and higher-order connections within the content.
KG-based recommendation systems aim to identify paths between items that users have interacted with and potential items to recommend by modeling higher-order relationships between users and items~\cite{ren2024explicit,sakurai2021listener,sakurai2021user,wang2024reinforced}.
Since KG-based recommendation systems combine collaborative and content-based filtering, they uncover latent relationships between users and items, enabling more accurate and explainable recommendations.

In recent years, Large Language Models (LLMs) have attracted attention as repositories rich in information learned from the web data in the recommendation fields.
Various recommendation approaches have been proposed that treat LLMs as recommendation systems due to their ability to generalize across diverse domains and understand complex user preferences~\cite{du2024enhancing,tan2024idgenrec,xu2024openp5,zhao2024let}. 
These approaches employ the meta-knowledge—or recommendation literacy—that LLMs have acquired, specifically understanding ``what kind of content users who have interacted with certain content might prefer." 
Such LLM-based recommendation approaches enable zero-shot recommendations without the need for training specific domain datasets.

Both KG-based and LLM-based recommendation models are intriguing approaches, however, each has significant issues. 
Firstly, KG-based recommendation models face two fundamental problems. 
One is that they often fail to consider temporal information. 
Most KG-based methods lack temporal data on edges or nodes, making it difficult to generate recommendations that adapt to changes in user preferences over time~\cite{dai2020survey}. 
The other is that they perform poorly under cold-start conditions where the target user has insufficient interaction data.
The performance of many collaborative filtering–based recommendation methods, including most KG-based ones, heavily depends on the user's interaction history~\cite{panda2022approaches}.
Secondly, LLM-based recommendation models encounter scalability challenges. 
Due to the limited number of tokens that can be input into the LLM, it is impractical to feed information of the entire dataset typical of recommendation domains, which includes massive amounts of user interactions and item metadata~\cite{hou2024large,wu2024survey}.
Consequently, accurately predicting ground truth items from a large pool of candidates using other users' interaction histories becomes difficult. 
In this way, KG-based and LLM-based recommendation models each have independent issues and are still underdeveloped research fields.

We argue that the weaknesses of KG-based and LLM-based recommendation models can be overcome by complementing each other's strengths. 
The weaknesses of KG-based recommendations can be mitigated by the capabilities of LLMs, such as prompt engineering and generative abilities.
For the first issue—lack of consideration for temporal information—we can function as a time-aware recommendation system by inputting prompts that specify temporal awareness. For the second issue—the cold-start problem—the superior generative abilities of LLMs allow us to generate textual representations of user preferences from a small amount of interaction data.
Conversely, we can overcome the weaknesses of LLM-based recommendations by leveraging the characteristic strength of KGs as large-scale databases with well-organized graph structures. 
KGs can store vast amounts of user interactions and content information. 
By allowing the LLM to determine a rough exploration strategy within the KG based on limited user input, we can circumvent scalability issues.
Thus, it is expected that effective recommendations that capitalize on their respective strengths by constructing an approach that bridges KGs and LLMs can be achieved.

In this paper, we propose an LLM's Intuition-aware Knowledge graph Reasoning model (LIKR) for cold-start sequential recommendation. 
Our main idea is to treat the LLM as a reasoner that outputs intuitive exploration strategies for the KG. 
Inspired by~\cite{hou2024large}, we incorporate temporal awareness through prompt engineering and generate textual representations of user preferences from limited interactions, allowing LIKR to improve recommendation performance in cold-start scenarios.
By representing the target dataset as a KG and constraining the LLM's output to KG exploration strategies, we can avoid scalability issues. 
Furthermore, we train a KG path reasoning agent using reinforcement learning, integrating the LLM's text-based output with existing recommendation metrics. 
We introduce a reinforcement learning algorithm that directly learns the agent's policy from scalar rewards, enabling us to combine multiple recommendation strategies—including user preferences predicted by the LLM and KG embedding–based metrics.
Our reinforcement learning agent explores the KG and determines recommended items based on a carefully designed reward function that considers these three metrics. 
By combining the general knowledge and generative capabilities of the LLM with the domain-specific insights of the KG, our model achieves effective sequential recommendation performance.

The contributions of our proposed method are threefold.
\begin{itemize}
      \item To the best of our knowledge, LIKR is the first approach to using LLM as a KG path reasoner for recommendation.
      \item To integrate different types of recommendation strategies, we introduced reinforcement learning and designed a reward function that incorporates the advantages of both LLMs and KGs.
      \item Through experiments on real-world datasets, we demonstrate that our model outperforms existing state-of-the-art methods in cold-start sequential recommendation scenarios.
\end{itemize}

\section{Related Work}
\label{RW}
\subsection{Knowledge Graph-based Recommendation}
KG-based recommendation systems aim to provide more advanced recommendations by modeling rich relational information about items and users, leveraging both collaborative filtering and content-based characteristics~\cite{ji2021survey,wang2021learning}.
Several approaches use paths that exist between entities in the KG, making recommendations based on information obtained through these paths~\cite{geng2022path,song2019ekar,wang2022multi}. 
These approaches, known as path-based methods, allow for multi-step capture of relationships between users and items and incorporate richer contextual information~\cite{balloccu2023knowledge,guo2020survey}. 
Xian et al.~\cite{xian2019reinforcement} introduced PGPR, which employs reinforcement learning to uncover multi-hop paths between users and items, subsequently offering recommendations aligned with these reasoning paths. 
In another study, Xian et al.~\cite{xian2020cafe} presented CAFE, a method that devises a user profile adept at capturing user-centric patterns, streamlining the recommendation process through efficient path reasoning. 
Tai et al.~\cite{tai2021user} proposed UCPR, which integrates the concept of a user's demand portfolio, refining the adaptability and efficiency of the path selection mechanism. 
These path-based methods can be effective even when the KG is large and complex, by applying techniques that efficiently search and evaluate paths.

However, these KG-based recommendations still face the cold-start problem because their accuracy relies on collaborative filtering~\cite{panda2022approaches}. 
Moreover, since most methods do not consider temporal information, it is hard to make recommendations that adapt to changes in user preferences. 
Some methods~\cite{chen2021temporal,zhao2022time} address the lack of temporal consideration by dynamically generating the KG; however, these approaches demand considerable effort.
Our approach addresses these issues with less effort by leveraging the flexible prompt engineering of LLMs and their ability to generate content from limited inputs.

\subsection{Large Language Model-based Recommendation}
The emergence of LLMs~\cite{bai2022constitutional,chowdhery2023palm,radford2018improving,touvron2023llama} has significantly impacted the field of Natural Language Processing (NLP), demonstrating substantial reasoning abilities and comprehensive world knowledge. 
Recently, various approaches have been proposed that treat LLMs themselves as recommendation systems. 
These methods learn based on the knowledge obtained during the pre-training of LLMs without using fine-tuning specific training data. 
Early study~\cite{sileo2022zero} showed that using GPT-2 directly as a recommendation model could compete with supervised matrix factorization when the number of users is very small. 
Hou et al.~\cite{hou2024large} enhanced the zero-shot recommendation capabilities of GPT-3.5 and GPT-4 by designing prompts that make the order of interaction histories explicit and eliminate popularity and position biases.
However, due to the limited number of tokens that can be input, it is challenging for these approaches to process the entire dataset in the general recommendation domains, which includes large amounts of user interactions and item metadata. 
Furthermore, there is a scalability issue because the number of candidate items for recommendation is limited. 
By treating the KG as a database that holds the dataset, our approach addresses the scalability problem.

Recently, approaches that employ both KGs and LLMs jointly have also been proposed. 
Yang et al.~\cite{yang2024common} proposed CSRec, which constructs a KG from commonsense extracted from LLMs and uses it in the knowledge-aware recommendation. 
In another study, Yang et al.~\cite{yang2024sequential} proposed LRD, which uses LLMs as item embedding extractors and integrated them into recommendations based on KG embeddings.
Different from these approaches, our LIKR uses LLM as a reasoner to determine the exploration strategy for the KG.
LIKR can collaboratively leverage the recommendation literacy held by the LLM and the semantic relationships maintained by the KG.

\section{LIKR: LLM's Intuition-aware Knowledge Graph Reasoning Model}
\label{PM}
\begin{figure}[t]
    \centering
    \includegraphics[width=1.00\columnwidth]{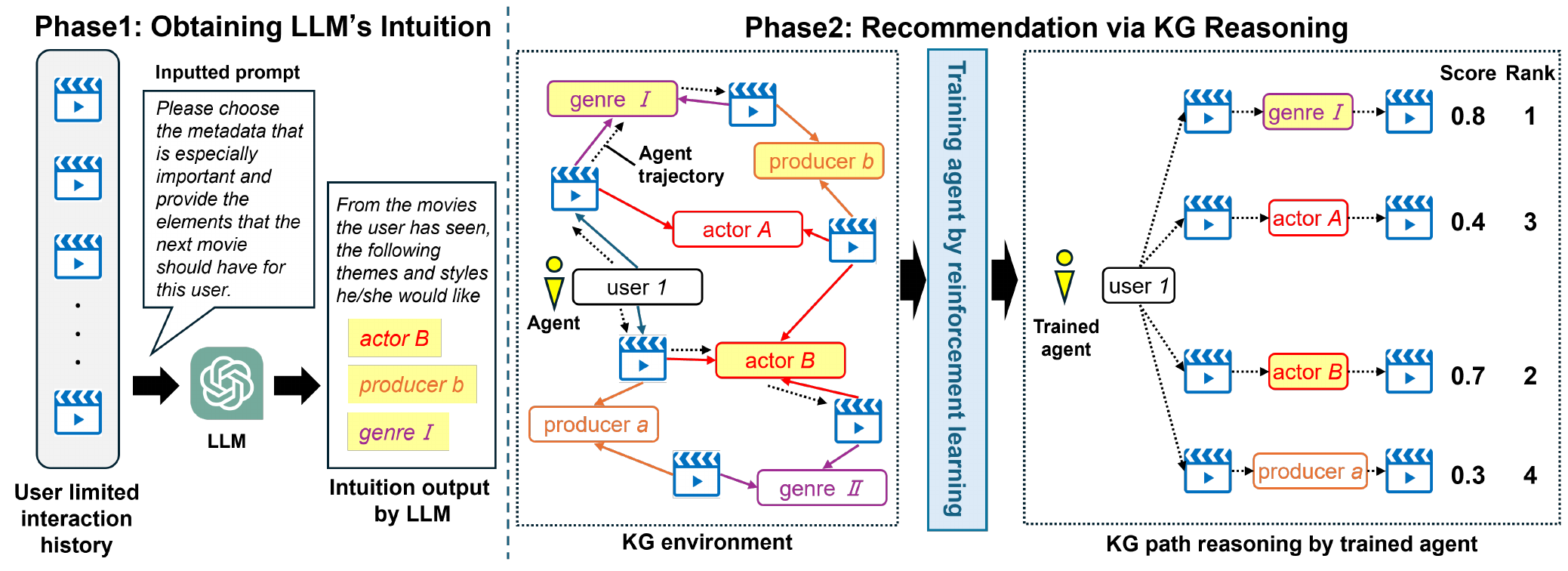}
    \caption{An overview of the proposed LIKR framework.}
    \label{fig:overview}
\end{figure}

In this section, we provide a detailed explanation of LIKR. 
We define the cold start setting as a scenario where the user's interaction history is sparse. 
In the cold start sequential recommendation task, the system predicts the user's next interaction items despite the limited interaction data. 
LIKR addresses this challenge in two phases: (1) obtaining the LLM's intuition about the user’s future preferences, and (2) making recommendations via KG reasoning based on the obtained LLM’s intuitions and reinforcement learning. 
An overview of the LIKR framework is shown in Fig.~\ref{fig:overview}.

\subsection{Obtaining LLM's Intuition about User's Future Preference}
\label{LLMInt}
In the first phase, we obtain LLM's intuition about a user's future preferences.
The obtained LLM's intuition affects the determination of the reasoning strategy for the KG described in Section~\ref{KGreasoning}.
Following the approach in~\cite{hou2024large}, we treat the LLM as a reasoner, using prompts that consider temporal awareness to predict future preferences.

We input a temporally-aware prompt into the LLM, which includes the limited interaction history ${\it {\mathcal H}_{u}}=\{i^{u}_{1}, i^{u}_{2}, ..., i^{u}_{N^{u}_{i}}\}$ (in chronological order of interaction time) of the user ${\it u}$ and the set of metadata types ${\mathcal Y}^{d}=\{y^{d}_{1}, y^{d}_{2}, ..., y^{d}_{N^{d}_{y}}\}$ in the recommendation dataset of the domain ${d}$.
Here, ${i^{u}}$ denotes an item with which user ${u}$ has interacted, $N^{u}_{i}$ denotes the total number of items user ${u}$ has interacted with.
Also, ${y^{d}}$ denotes the one of metadata types included in the recommendation dataset of the domain ${d}$, and $N^{d}_{y}$ denotes the total number of metadata types in the recommendation dataset of the domain ${d}$, respectively.
For example, in the dataset of the movie domain (${d}=``{\rm movie}"$), the inputted prompt is set as follows: 
\begin{center}
\begin{minipage}[t]{0.95\textwidth} 
{\it ``I am thinking of recommending the \uline{next} movie for a user to watch. The user has watched the following movies \uline{in this order, with the more recently watched movies appearing at the end}:
${i^{u}_{1}}$, ${i^{u}_{2}}$, ..., ${i^{u}_{N^{u}_{i}}}$.\\
Based on the metadata types such as $y^{\rm movie}_{1}$, $y^{\rm movie}_{2}$, ..., $y^{\rm movie}_{N^{\rm movie}_{y}}$, please choose the metadata that is especially important, and provide the elements that the \uline{next} movie should have for this user.\\
Please format the output as follows:\\
$y^{\rm movie}$: element"},
\end{minipage}
\end{center}
where the \uline{underlined parts} indicate the temporally-aware parts.
By providing the prompt to the LLM, we can obtain the set of metadata elements ${\mathcal E}_{u}$.
${\mathcal E}_{u}$ corresponds to the ``{\it element}" in the above prompt.
LLMs leverage the transformer architecture, specifically the self-attention mechanism, which allows simultaneous attention to all parts of the input sequence. This architecture enables effective interpretation of both the user's interaction history and the associated metadata while accounting for temporal flow. The ``intuition" derived from the LLM extends beyond basic pattern recognition, utilizing the model's capacity to contextualize user behavior based on its pre-trained knowledge and trends.
The advantage of this approach lies in its adaptability and scalability. 
LLMs can utilize pre-trained data across various domains, allowing interpretation of interaction sequences and metadata types without extensive domain-specific fine-tuning.
The obtained metadata element set ${\mathcal E}_{u}$ reflects the user's predicted future preferences and serves as a foundational step for refining the KG reasoning strategy.

\subsection{Recommendation via KG Reasoning Based on LLM’s Intuition and Reinforcement Learning}
\label{KGreasoning}
In the second phase, we determine the items to be recommended by conducting KG reasoning using the LLM's intuition obtained in Section~\ref{LLMInt}.
First, we construct a KG ${\mathcal G} = ({\mathcal V}, {\mathcal R})$ from the dataset.
Here, ${\mathcal V}$ is the set of entities, and ${\mathcal R}$ represents the relations between these entities. 
The entities consist of the set of users, items, and metadata of each item.
To leverage semantic information of the constructed KG and capture the diversity of relationships, we embed the KG.
For KG embedding, we employ the TransE~\cite{bordes2013translating} method. In TransE, entities and relations are embedded into the same space, and relations are modeled as vector translations between entities. Specifically, the relationship between two entities is represented by adding their embedding vectors, where the relation $r$ moves entity $e_1$ to $e_2$, as defined by the following equation:
\begin{equation}
\mathbf{e}_1 + \mathbf{r} \approx \mathbf{e}_2,
\end{equation}
where $\mathbf{e}_1$ and $\mathbf{e}_2$ are the embedding vectors of entities $e_1$ and $e_2$, respectively, and $\mathbf{r}$ is the embedding vector of relation $r$. The model learns the embedding vectors of entities and relations to satisfy this equation.
Due to its simple structure, TransE is computationally efficient, and it has been experimentally shown to roughly capture many-to-many relationships in the context of the large-scale graph data handled in the recommendation system domain~\cite{xian2019reinforcement,yang2023content}.

To achieve a recommendation strategy that combines both LLM-based intuition and KG embedding-based methods, we employ KG reasoning via reinforcement learning. Reinforcement learning enables us to solve recommendation tasks by leveraging multiple metrics cooperatively, through the definition of a reward function. Specifically, following the framework of \cite{xian2019reinforcement}, we model the recommendation task as a Markov Decision Process (MDP), where the KG reasoning component is central.
The MDP is defined as follows.
\begin{itemize}
    \item \textbf{State}:  
    At each time step $t$ ( $t = 0, 1, \ldots, T$; $T$ being the total number of steps in the exploration), the state is represented as $s_t = (u, e_t, h_t)$, where $u \in \mathcal{U}$ is the user entity, $e_t$ is the entity reached at step $t$, and $h_t$ is the history of previously encountered entities and relations. Here, ${\mathcal U}$ denotes the set of users.
    The initial state is $s_0 = (u, u, \varnothing)$.

    \item \textbf{Action}:  
    At each time step $t$, the agent selects an action $a_t$ from the set of valid outgoing edges of $e_t$, transitioning the agent to a new state $s_{t+1}$. To encourage diverse exploration, the action space $A(s_t)$ excludes entities and relations that have already been visited.

    \item \textbf{Reward}:  
    The reward $R_t$ is computed based on the score of the path from the user node $e_0$ to the item node $e_k$. It integrates two key metrics from our recommendation strategies: one derived from LLM's intuition and the other from KG embeddings.
    \begin{itemize}
        \item \textbf{LLM intuition-based reward $R^{\rm LLM}_t$}:  
        This reward is given when the agent reaches a node included in the user’s metadata set $\mathcal{E}_u$. Formally, it can be defined as
        \begin{equation}
        R^{\rm LLM}_t = 
        \begin{cases}
        1, & \text{if} \ e_0 = u \ \text{and} \ e_t \in \mathcal{E}_u \\
        0, & \text{otherwise}.
        \end{cases}
        \end{equation}
        \item \textbf{KG embedding-based reward $R^{\rm KG}_t$}:  
        This reward is calculated based on the inner product of the embedding vectors of entities along the path. Using KG embeddings from TransE, it can be defined as
        \begin{equation}
        R^{\rm KG}_t = \sum_{i=0}^{t} \langle \mathbf{e}_i, \mathbf{e}_{i+1} \rangle,
        \end{equation}
        where $\mathbf{e}_i$ represents the embedding of entity $e_i$.
    \end{itemize}
    The overall reward ${R_t}$ at time step $t$ is
    \begin{equation}
    R_t = R^{\rm LLM}_t + \alpha R^{\rm KG}_t,
    \end{equation}
    where $\alpha$ is a hyperparameter that balances the contribution of the two strategies.
\end{itemize}

Based on the above MDP setup, we optimize the policy network using the REINFORCE algorithm \cite{williams1992simple}. 
Finally, we use the agent with the learned policy network to determine the recommended items through KG reasoning.
Starting from a user node $u \in {\mathcal U}$, the valid actions (edges) from the current state $s_t$ are evaluated based on the learned policy.
To perform efficient path reasoning, we employ beam search. 
At each step $t$, the top $z$ actions are selected based on the policy score, and this process is repeated recursively $T$ times.
All generated paths are stored and used to produce the final Top-{\it N} recommendations.

\section{Experiments}
\label{Ex}
To answer the following research questions (RQs), we conduct experiments on real-world knowledge-based recommendation datasets.

\noindent
\textbf{RQ1:} 
How does LIKR perform overall in comparison to conventional methods in cold-start sequential recommendation scenarios?

\noindent
\textbf{RQ2:}
How do the types of LLMs and the prompts affect the recommendation performance of LIKR?

\noindent
\textbf{RQ3:} 
How do the different recommendation strategies included in the reward function impact the recommendation performance of LIKR?

\subsection{Experimental Setup}
\textbf{Datasets:} 
We used two representative real-world datasets in knowledge-aware recommendation: MovieLens-1M~\cite{harper2015movielens} and Lastfm-1M~\cite{balloccu2023knowledge}. 
The statistics of the datasets are shown in Table~\ref{dataset_statistics}.
Lastfm-1M is provided by~\cite{balloccu2023knowledge} and has been adjusted from Lastfm-1B~\cite{schedl2016lfm} to enable fair comparisons in KG-based recommendations.
Our approach requires the text of item titles to obtain outputs from the LLM and the text of item metadata to connect the LLM to the KG. 
Therefore, we excluded other representative datasets where items and metadata are only available as IDs and cannot be utilized. 
Following a time-based holdout strategy, we split the data into 60\% for the training set, 20\% for the validation set, and 20\% for the test set, in order of the oldest interactions.
Furthermore, to handle cold-start sequential recommendation scenarios in our experiments, we limited the number of user-item interactions in the training data. 
Following ~\cite{cai2022exploring,kang2018self,pan2020rethinking,sakurai2022deep}, we retained only the most recent 10 user-item interactions as training data and deleted all earlier interactions.
Focusing on the latest interactions allows us to evaluate the predictive performance of users' subsequent interests and preferences. 

\begin{table}[t]
\centering
\caption{Dataset statistics.}
\scalebox{0.9}[0.9]{
\begin{tabular}{llcc}
\hline
& & MovieLens-1M & Lastfm-1M \\
\hline
\multirow{4}{*}{\textbf{Interactions}} & Users & 6,040 & 4,817 \\
 &  Items & 2,984 & 12,492 \\
 &  Interactions & 932,295 & 1,091,275 \\
 &  Density & 0.05 & 0.01 \\
\hline
\multirow{3}{*}{\textbf{KG}} &  Entities (Types) & 13,804 (12) & 17,492 (5) \\
 &  Relations (Types) & 193,089 (11) & 219,089 (4) \\
 &  Sparsity & 0.0060 & 0.0035 \\
\hline
\end{tabular}
}
\label{dataset_statistics}
\end{table}

\noindent
\textbf{Comparison Methods:} 
We compared our proposed LIKR with recommendation methods across two categories: sequential models and KG-based models.
For sequential models, we employed GRU4Rec~\cite{hidasi2015session}, which leverages Gated Recurrent Units (GRUs)~\cite{cho2014learning} to capture sequential dependencies in session-based recommendations, and SASRec~\cite{kang2018self}, which uses self-attention to model user behavior for next-item recommendations.
In the experiments in~\cite{hidasi2015session,kang2018self}, the validation data is included during testing. 
On the other hand, KG-based methods explore the KG graph using only the training data as a clue. 
In our experiments, from the perspective of fairness, we follow the KG-based evaluation method and do not input the validation data during testing.
For KG-based models, we employed three representative path-based models: PGPR~\cite{xian2019reinforcement}, CAFE~\cite{xian2020cafe}, and UCPR~\cite{tai2021user}.
The sequential methods were implemented using RecBole~\cite{recbole}, a recommendation research library, while the KG-based methods followed the implementation of~\cite{balloccu2023knowledge}.

\noindent
\textbf{Evaluation Metrics:}
We used recall@{\it k} and normalized Discounted Cumulative Gain (nDCG)@{\it k} \{{\it k}= 20, 40\}, which are standard metrics for measuring top-{\it k} recommendation performance in knowledge-aware and sequential recommendation.

We utilized the metadata sets provided with each dataset as input to the LLM. 
However, ``wikipage" and ``category" from the MovieLens-1M dataset and ``genre" from the Lastfm-1M dataset were excluded due to the difficulty in matching the LLM output with the metadata IDs of each dataset.
Based on the assumption that shorter paths are more reliable when users interpret recommendation reasons, we set the maximum path length to 3.
The embedding size of the KG using TransE was set to 100.
The maximum size of the action space was set to 400.
To promote path diversity, we applied action dropout at a rate of 0.5 to the pruned action space.
The discount factor of reinforcement learning was set to 0.99.
The hyperparameter $\alpha$ in the reward function was tuned from \{0, 0.25, 0.5, 0.75, 1\}.
For all datasets, the model was trained for 100 epochs using the Adam optimizer, with a learning rate of 0.001 and a batch size of 64.
The sampling sizes for the beam search were set to 4 for the first hop, 2 for the second hop, and 1 for the third hop in the MovieLens-1M dataset, and 10 for the first hop, 5 for the second hop, and 1 for the third hop in the Lastfm-1M dataset.

\subsection{Overall Performance Comparison (\textbf{RQ1})}
To answer RQ1, we compared the recommendation performance of LIKR with that of both sequential and KG-based recommendation models. 
The results are shown in Table~\ref{performance_comparison}. 
In this experiment, we employed GPT-4o-preview, the latest GPT-based model as of October 2024.
When comparing sequential recommendation models with LIKR, LIKR consistently outperforms all evaluation metrics and datasets.
GRU4Rec and SASRec, designed for session-based recommendations, demonstrate competitive or even superior performance to KG-based models in cold-start sequential recommendation scenarios. 
The lower nDCG values for sequential models may be attributed to their design, which optimizes for predicting a single item using both training and validation data as input. 
By incorporating the LLM’s temporal awareness, LIKR achieves higher recommendation performance in cold-start sequential recommendation scenarios than in sequential models.

Next, when compared with KG-based recommendation models, LIKR outperforms all other comparison methods in all metrics except metrics@40 for Lastfm-1M. 
In particular, LIKR surpasses PGPR, a method based on path reasoning using reinforcement learning, in all evaluation metrics. 
This demonstrates the effectiveness of integrating the LLM's intuition into reinforcement learning. 
The fact that LIKR is inferior to UCPR on the Lastfm-1m metrics@40 suggests that, while LIKR can predict short-term preferences with relatively high accuracy since it takes time series into account, it has limitations when it comes to predicting preferences in the distant future.
Overall, LIKR shows higher values than the representative recommendation models for the two categories, indicating that it can deliver good recommendation performance in the cold-start scenario.

\begin{table}[t]
\centering
\caption{Comparison of Performance Metrics on MovieLens-1M and Lastfm-1M. The highest value is shown in \textbf{bold}, and the second highest value is shown in \underline{underlined text}.}
\scalebox{0.8}[0.8]{
\begin{tabular}{l|cc|cc|cc|cc}
\hline
\multirow{3}{*}{Method} & \multicolumn{4}{c|}{MovieLens-1M} & \multicolumn{4}{c}{Lastfm-1M} \\
\cline{2-9}
 & \multicolumn{2}{c|}{ Metrics@20 [\%] } & \multicolumn{2}{c|}{Metrics@40 [\%]} & \multicolumn{2}{c|}{Metrics@20 [\%]} & \multicolumn{2}{c}{Metrics@40 [\%]} \\
\cline{2-9}
 & Recall & nDCG & Recall & nDCG & Recall & nDCG & Recall & nDCG\\
\hline
GRU4Rec~\cite{hidasi2015session} & \underline{4.77} & 4.73 & 9.02 & 6.30 & 1.22 & 2.65 & 2.16 & 2.67 \\
SASRec~\cite{kang2018self} & 4.63 & 4.47 & 8.72 & 6.04 & 0.77 & 1.28 & 1.27 & 1.36 \\\hline
PGPR~\cite{xian2019reinforcement} & 4.63 & \underline{18.53} & \underline{9.32} & 23.12 & 1.27 & \underline{12.73} & 2.29 & 14.30 \\
CAFE~\cite{xian2020cafe} & 4.14 & 15.35 & 7.27 & 22.86 & 1.19 & 10.10 & 2.20 & 15.05 \\ 
UCPR~\cite{tai2021user} & 2.64 & 14.47 & 7.31 & \underline{23.28} & \underline{1.38} & 12.34 & \textbf{3.32} & \textbf{18.72} \\\hline
\textbf{LIKR (GPT-o1-preview)} & \textbf{4.83} & \textbf{19.14} & \textbf{9.46} & \textbf{23.80}  & \textbf{1.42} & \textbf{12.93} & \underline{2.83} & \underline{16.32} \\
\hline
\end{tabular}
}
\label{performance_comparison}
\end{table}

\subsection{Impact Analysis of LLM Types and Prompts (\textbf{RQ2})}
To answer RQ2, we compared the recommendation performance by varying the LLM and prompt used in LIKR.
First, to assess the effect of the LLM type on LIKR's performance, we used six representative LLMs: Gemini1.5-flash, Claude3.5-sonnet, GPT-3.5-turbo, GPT-4-turbo, GPT-4, and GPT-o1-preview. 
The results are shown in Table~\ref{LLM_comparison}. 
In both datasets, LIKR achieves high recommendation performance overall when GPT-o1-preview is used. GPT-o1-preview can generate refined outputs that more accurately reflect the prompt's content by examining the inference process in detail. 
Therefore, by considering the interaction history and time series input to the LLM, GPT-o1-preview can predict future user preferences precisely. 
As we move to newer models in the GPT series, the recommendation performance improves; however, the difference between GPT-3.5-turbo and GPT-4 is not significant. 
Gemini1.5-flash does not perform as well, likely because it does not understand the output format specified in the prompt and more frequently produces outputs that reject preference prediction compared to other models.

Next, we compared the recommendation performance of LIKR using GPT models with prompts that either consider or ignore temporal awareness. 
The prompts that ignore the temporal awareness omit the underlined portion of the prompt shown in Section~\ref{LLMInt}. 
The results are shown in Table~\ref{TA_comparison}. 
For all metrics except nDCG@20 in GPT-4, incorporating temporal awareness into the prompts improves recommendation performance. 
In particular, GPT-o1-preview is sensitive to the prompt content, and its outputs often vary depending on whether the temporal awareness is included in the prompt. 
Additionally, the recall and nDCG of some LIKR models that ignore temporal awareness are lower than those of the conventional sequential recommendation model. 
In this way, the choice of LLM and the prompt significantly impact the effectiveness of the proposed method.

\begin{table}[t]
\centering
\caption{Comparison of LIKR performance when using different LLMs.}
\scalebox{0.93}[0.93]{
\begin{tabular}{l|cc|cc}
\hline
\multirow{3}{*}{Method} & \multicolumn{2}{c|}{MovieLens-1M} & \multicolumn{2}{c}{Lastfm-1M} \\
\cline{2-5}
 & \multicolumn{2}{c|}{Metrics@20 [\%]} & \multicolumn{2}{c}{Metrics@20 [\%]} \\
\cline{2-5}
 & Recall & nDCG & Recall & nDCG\\
\hline
LIKR (Gemini1.5-flash) & 4.61 & 18.90 & 1.13 & 11.41 \\
LIKR (Claude3.5-sonnet) & 4.76 & \textbf{19.16} & 1.30 & \underline{12.79} \\
LIKR (GPT-3.5-turbo) & 4.73 & 19.01 & 1.29 & 12.33 \\
LIKR (GPT-4-turbo) & 4.76 & 18.92 & 1.32 & 12.33 \\
LIKR (GPT-4) & \underline{4.78} & 19.05 & \textbf{1.36} & 11.56 \\
LIKR (GPT-o1-preview) & \textbf{4.83} & \underline{19.14} & \underline{1.35} & \textbf{12.93} \\
\hline
\end{tabular}
}
\label{LLM_comparison}
\end{table}

\begin{table}[t]
\centering
\caption{Comparison of LIKR performance when using different prompts.}
\scalebox{0.93}[0.93]{
\begin{tabular}{l|cc|cc}
\hline
\multirow{3}{*}{Method} & \multicolumn{2}{c|}{MovieLens-1M} & \multicolumn{2}{c}{Lastfm-1M} \\
\cline{2-5}
 & \multicolumn{2}{c|}{Metrics@20 [\%]} & \multicolumn{2}{c}{Metrics@20 [\%]} \\
\cline{2-5}
 & Recall & nDCG & Recall & nDCG\\
\hline
LIKR (GPT-4-turbo w/o temporal awareness) & 4.62 & 18.78 & 1.20 & 12.30 \\
LIKR (GPT-4-turbo) & \textbf{4.76} & \textbf{18.92} & \textbf{1.32} & \textbf{12.33} \\\hline
LIKR (GPT-4 w/o temporal awareness) & 4.66 & 18.97 & 1.28 & \textbf{11.89} \\
LIKR (GPT-4) & \textbf{4.78} &  \textbf{19.05} & \textbf{1.36} & 11.56 \\\hline
LIKR (GPT-o1-preview w/o temporal awareness) & 4.65 & 18.60 & 1.36 & 12.65 \\ 
LIKR (GPT-o1-preview) & \textbf{4.78} & \textbf{19.14} & \textbf{1.42} & \textbf{12.93} \\
\hline
\end{tabular}
}
\label{TA_comparison}
\end{table}

\subsection{Impact Analysis of Recommendation Strategies (\textbf{RQ3})}
\begin{figure}[t]
    \centering
    \includegraphics[width=1.00\columnwidth]{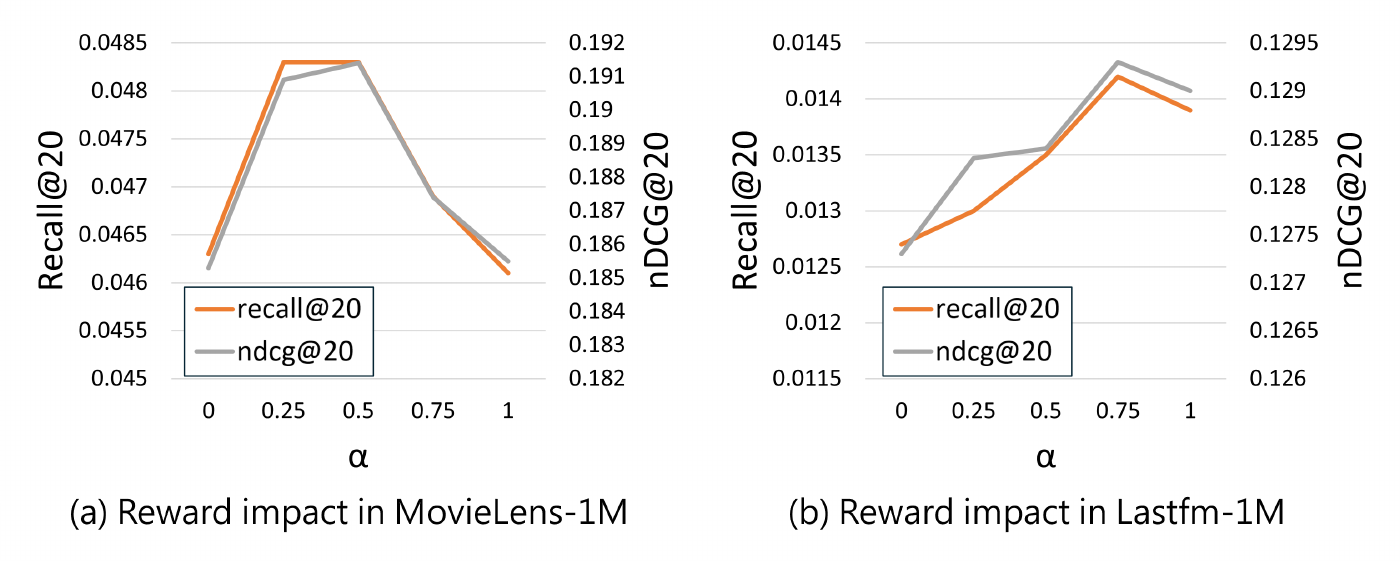}
    \caption{Impact of LLM's intuition-based and KG embedding-based rewards.}
    \label{fig:reward}
\end{figure}

To answer RQ3, we evaluated the recommendation performance of LIKR by varying the hyperparameter $\alpha$ in the reward function.
The results are shown in Fig.~\ref{fig:reward}.
From Fig.~\ref{fig:reward}~(a), we observe that in the MovieLens-1M dataset, the highest performance is achieved when $\alpha$ is between 0.25 and 0.5. 
From Fig.~\ref{fig:reward}~(b), we see that in the Lastfm-1M dataset, the highest performance occurs when $\alpha$ is 0.75.
This suggests that the optimal proportion of LLM's intuition used differs between the movie and music domains.
Additionally, when $\alpha$ is 0 or 1, the recommendation performance decreases in both datasets, indicating that it is important to use both the KG embeddings and LLM's intuition collaboratively in each dataset.
In this way, it was shown that the two types of reward functions each have a significant impact on recommendation performance.

\section{Conclusion}
\label{Conc}
In this paper, we have proposed LIKR, a recommendation model based on reinforcement learning and KGs that leverage the intuition of LLMs.
Our proposed method treats the LLM as a reasoner that outputs intuitive exploration strategies over the KG, achieving improved recommendation performance in cold-start sequential recommendation scenarios.
We demonstrated the effectiveness of LIKR by comparing it with conventional recommendation models in two categories using two real-world datasets.
Our strategy for using the LLM is a simple and flexible idea that can overcome the scalability limitations of LLM-based recommendation systems and is expected to be applicable to reward functions and loss functions in various recommendation models.
In future work, we plan to research a recommendation model that leverages more sophisticated outputs of the LLM by fine-tuning the LLM itself for KG reasoning.

\bibliographystyle{splncs04}
\bibliography{refs}
\end{document}